\documentclass[11pt]{article}
\usepackage{geometry}                
\usepackage{graphicx}
\usepackage{amssymb}
\usepackage{epstopdf}
\usepackage{fancyhdr}

\title{Mathematical and Statistical Opportunities in Cyber Security\thanks{This work was supported by the Director, Office of Science, of the U.S. Department of Energy under Contract No. DE-AC02-05CH11231.} }
\author{Juan Meza\thanks{High Performance Computing Research, Lawrence Berkeley National Laboratory ({\tt JCMeza@lbl.gov}).} \and Scott Campbell\thanks{National Energy Research Scientific Computing Center ({\tt SCampbell@lbl.gov}).} \and David Bailey\thanks{High Performance Computing Research, Lawrence Berkeley National Laboratory ({\tt DHBailey@lbl.gov}).}}

\date{LBNL-1667E \\ March 2009 }                                           

\begin{document}

\maketitle
\subsection*{\centerline{Abstract}}
The role of mathematics in a complex system such as the Internet has yet to be deeply explored.    In this paper, we summarize some of the important and pressing problems in cyber security from the viewpoint of open science environments.  We start by posing the question ``What fundamental problems exist within cyber security research that can be helped by advanced mathematics and statistics?"  Our first and most important assumption is that access to real-world data is necessary to understand large and complex systems like the Internet.   Our second assumption is that many proposed cyber security solutions could critically damage both the openness and the productivity of scientific research.   After examining a range of cyber security problems, we come to the conclusion that the field of  cyber security poses a rich set of new and exciting research opportunities for the mathematical and statistical sciences.
\newpage

\section{Introduction}
A cyber security incident of some sort makes the news headlines on an almost daily basis.  The examples are numerous, from individual users information loss, to worms and computer viruses, to large scale criminal behavior precipitated by organized crime and nation states.   More recently~\cite{Markoff2009}, the large-scale use of botnets for distributing e-mail spam, distributed denial of service attacks, and distributing other malware has led to an informal alliance of computer security experts from across the world.
Not surprisingly, the rise in cyber security incidents is due in large part to the rise in the use of computers and the Internet in all areas of society.  In fact, according to~\cite{AlPaTe07}, {\em ``Incessant scanning of hosts by attackers looking for vulnerable servers has become a fact of Internet life"}.  Therefore it comes as no surprise that scientific research has also been affected by cyber security problems.  Indeed, because so much of science has embraced and come to rely so heavily on computing resources, it is particularly vulnerable to cyber security issues.  

While the growth in computational, network, and data resources has completely changed the way that basic scientific research is conducted, this has in general not been reflected in the way that computer science has addressed challenges facing cyber security research.  A recent report by the National Research Council~\cite{NAPCyber} states that, ``research can produce a better understanding of why cyberspace is as vulnerable as it is and that such research can lead to new technologies and policies and their effective implementation, making cyberspace safer and more secure."  However, the committee was also careful to mention that, ``there are no single or even small number of silver bullets that can solve the cybersecurity problem".  At almost the same time, a grass roots community effort recently released a report~\cite{Catlett2008} that outlines several areas for a science-based cyber security research program.  The report outlined three focus areas: predictive awareness for secure systems, self-protective data and software, and trustworthy systems from trustworthy components.  The three focus areas were further subdivided into specific research areas.   A similar report~\cite{Dunlavy2009} highlighted three major challenge areas: modeling large-scale networks, threat discovery, and network dynamics.

Interestingly, the role of mathematics in a complex system such as the Internet has yet to be deeply explored.   Willinger and Paxson~\cite{WiPa98} pointed out as early as 1998 that, ``the Internet is a new world, one where engineering wins out over tradition-conscious mathematics and requires {\em paradigm shifts} that favor a combination of mathematical {\em beauty} and high potential for contributing to pragmatic Internet engineering."  In that same spirit,  we would like to ask ``What fundamental problems exist within cyber security research that can be helped by advanced mathematics and statistics?"  In this paper, we summarize some of the more important and pressing problems in cyber security from the viewpoint of open science environments, and highlight those which we believe should be of interest to the general mathematical sciences community.  

We wish to stress the importance of using a science-based approach.  Our first and most important assumption is that, as in other scientific fields, access to real-world data is necessary to understand large and complex systems.  Applying mathematical models and advanced algorithms to this real-world data, it should be possible to develop {\em validated} predictive models that can then be used to develop more robust applications for workable cyber security.  With the rise of cyber security incidents and more persistent and stronger adversaries, it makes sense to move away from current ad-hoc, reactive methodologies and limited testing to more rigorous and repeatable approaches.  Our second assumption is that many proposed cyber security solutions could critically damage both the openness and the productivity of scientific research.  As such, we want to emphasize that using security models from other areas (e.g. the classified sectors) and relaxing the conditions usually results in a model that is inherently detrimental to open science.  We also note that many of the challenges in cyber security arise from characteristics of the Internet that make it difficult to model, such as: 1) the self-similar structure of network traffic, 2) the inherent dynamic nature of the Internet, and 3) the rapid growth in the Internet, both in terms of the number of components and size of the traffic.

This paper is divided into three sections: 1) data, 2) modeling, and 3) applications.  Section 2 make the case for good raw data as the foundation for basing cyber security models and policies.  Section 3 outlines some of the major areas where improved modeling is needed.  The final section describes the use of some models and data to implement applications that have been used to detect and deter adversaries. In all three areas, we highlight some of the mathematical opportunities that arise as researchers have studied these areas, including recent advances using statistical techniques. 
%
%
%
%
\section{The Importance of Data}

The need for trace and log data for scientific analysis is necessary not only to create accurate models, but to provide repeatability and verification of results.  In addition, ensuring that the data's integrity is maintained throughout the lifetime of its intended use is imperative in order to be able to validate the results of any scientific model.   As a fundamental building block of repeatable science, we see the lack of freely available raw data as an issue that is both critical for success and a problem that can be addressed through better mathematical modeling and techniques.     This section deals with three areas that lay the foundation for almost all of the cyber security defense mechanisms in place today: 1) the need for real-world data on which to base network models, 2) the need for developing methods that ensure data integrity, and 3) the need to handle large amounts of data in real or almost real time. 

\subsection{Data Anonymization and Cleaning}

The lack of public data sets for network modeling has been identified as a key weakness in current networking research~\cite{Paxson2004}.  In addition, most intrusion detection systems are based on anomaly detection for which one of the key assumptions is the availability of good training data.  The generally accepted assumptions for the training data include: 1) attack-free data is available, 2) simulated data is representative, and 3) network traffic is static.  Gates and Taylor~\cite{GaTa07}, however, challenge these assumptions and argue that most of these assumptions may not hold in many situations.    

Many sites are reluctant to publicly release network data for a variety of reasons including confidentiality, privacy, and security issues. Given the need for high-quality data however, some researchers have studied the question of how best to anonymize or sanitize the data so that it can be released publicly.    In all of these cases, there is an inherent tradeoff between the need to ensure security and privacy and the need for high-quality data that still represents real-world traffic data. This is known as the utility versus security trade-off~\cite{Slagell2006}.  Several approaches for sanitizing data have been suggested with various degrees of success~\cite{Bishopet06}.  It is also important to consider how effective a particular anonymization policy is and there have been some efforts to measure this~\cite{Lakkaraju2008}.  The diversity of techniques used in tackling the data indicate that a more systematic approach might be applied to this problem as well as methods for maximizing a particular utility function.  The application of data anonymization methods is not restricted to raw data however.  Higher layer abstractions such as NetFlow records (Cisco IOS NetFlow.  http://www.cisco.com/go/netflow) or Bro connection logs~\cite{Paxson98} are also extremely powerful for large scale measurement and modeling. 

In addition to releasing (possibly anonymized) real data, there are other ways of generating test data -- synthesis and reference data.  However, generating synthetic data can be problematic in terms of research value~\cite{McHugh2000}, while reference data (data recorded at locations such as honey pots which have no privacy constraints) may not fulfill the needs of the researcher due to the specific nature of the traffic~\cite{Bishopet06}.


\subsection{Data Integrity}
As the data sets in science grow and as our dependence on them for understanding science increases, there is a critical need for ensuring that data maintains its integrity over the lifetime of its intended use.  By integrity here, we mean the trustworthiness of either the data or resources and includes {\em data} integrity (content) and {\em origin} integrity (the source of the data).  As described by Bishop\cite{Bishop03}, integrity mechanisms fall into two main classes: prevention mechanisms and detection mechanisms.  New methods will need to be developed that can ensure that the data being generated by large experimental facilities such as the Large Hadron Collider, ITER, or any of the large accelerators, maintains its integrity during the course of the analysis of said data.  Likewise, as programs such as the Department of Energy's Scientific Discovery through Advanced Computing (SciDAC) expand towards exascale facilities, the data sets generated by the codes that the SciDAC program support can be expected to also grow in size.  This simulation output will also require methods for maintaining data integrity. One interesting new approach for securing the provenance of data was suggested by~\cite{Braun2008}.  They note that provenance can be modeled as a causality graph with annotations, where the causality graph describes the process that produced the data's present state.  As such, the graph can be represented as an immutable directed acyclic graph (DAG).  Although, they present a security solution, they also note that more research is required to construct a security model for causal graphs.  

\subsection{Real-time data}
In order to respond quickly to a cyber security attack, organizations need to analyze high-volumes of traffic data and detect anomalies in real time.  Dreger et al.~\cite{Dreger2004} cite some examples from two operational cases that consisted of networks with tens of thousands of hosts, transferring 2-3 terabytes of data/day and 44,000 packets/sec on average.  Some interesting work using statistical techniques such as sequential hypothesis testing has shown that this is possible~\cite{Jung2004}. The basic idea is to model accesses to local IP addresses as a random walk on one of two stochastic processes, corresponding respectively to a benign and a malicious process.  The use of sequential hypothesis testing is intriguing because it can be used to establish mathematical bounds on the expected performance of the algorithm.   While this work is quite promising, the authors point out that their work only focused on the detection of an attack from a {\em single} remote address.   New mathematical models will be needed to determine whether a coordinated attack from a set of remote addresses is taking place.   In addition, as networks increase in bandwidth and the number of hosts increase, it is clear that the data sets that need to be analyzed will also grow, requiring new mathematical methods that can scale with the traffic data.

%
%

\section{Modeling}
\subsection{Internet Modeling}
Modeling the Internet is well-known to be a difficult problem~\cite{Floyd2001,Paxson1997a}.  Some of the difficulties include the immense heterogeneity of the Internet and the rapid changes over time.  Floyd and Paxson~\cite{Floyd2001} proposed two strategies for developing meaningful simulations in the face of these difficulties: searching for invariants and judiciously exploring the simulation parameter space.  In terms of invariants, Floyd and Paxson suggest among others: diurnal patterns of activity, self-similarity, heavy-tailed distributions, and log-normal connection sizes.  Searching for these invariants can then be viewed as a problem in estimating the correct set of parameters from the data.  The second strategy proposed, that of judiciously exploring the parameter space was proposed as a way to cope with the heterogeneity and change in the Internet architecture.  Exploring the simulation parameter space can also be framed as a mathematical problem. For example, one way is to pose this as a problem in the design and analysis of computer experiments, for which there is already a great deal of literature.  As the number of parameters increase though, this approach can be quite limiting. Therefore, techniques for determining which parameters are the most important for a particular model or determining the sensitivity of the simulation to certain parameters will need to be developed.  Finally, one can also view this as an optimization problem in which one seeks to minimize certain behavior as defined by an appropriate cost function.

\subsection{Statistical Models}
Network traffic is often modeled as a Poisson process because it has good theoretical properties.  Internet traffic, however, has been shown to have some very complex statistical properties~\cite{Cleveland2000,Raftery2001}.  In fact, many studies have shown that simple Poisson models do not hold for real network traffic including both local area and wide area network traffic.  Several studies~\cite{Leland1994,Willinger1997a}  have shown instead that local area network can be modeled much better as a self-similar process.  An interesting study on Internet traffic data that describes these phenomena was provided by Cleveland and Sun~\cite{Cleveland2000}.  In addition to an excellent description of the problem, Cleveland and Sun suggest several challenges for handling traffic data including: statistical tools and models for point processes, marked point processes, and time series that account for nonstationarity, persistence, and distributions with long upper tails, 2) theoretical and empirical exploitations of the superposition of Internet traffic, and 3) integration of statistical models with network simulators.  An excellent bibliography on self-similar traffic modeling and analysis can be found in~\cite{WTE96}.  

Paxson and Floyd~\cite{Paxson1995} suggest that the Poisson-based models should also be abandoned for wide-area traffic.   In his studies of data sets from the teletraffic industry, Resnick~\cite{Resnick97} noted that traffic data often exhibit many non-standard characteristics such as heavy tails and long range dependence. He also described several estimation methods for the analysis of heavy tailed time series including  parameter estimation and model identification methods for autoregressions and moving averages. However, in the discussion that followed Resnick's paper, Willinger and Paxson~\cite{Willinger1997}  argued persuasively for using structural models that take into account the context in which the data arose as opposed to the black box modeling approach that is more commonly used.  Clearly there is a need for further research and development of statistical techniques and methods for  effectively  handling phenomena such as heavy tails and long-range dependence that arise in cyber security data.
 


%
%
%
%

\section{Network Intrusion and Attack Response}
Network intrusion and attack detection typically results in the application of a successful model built from real world data.  For example if one has a working model of how an external adversary might scan one or more hosts on a local network, one can build a simple detector based on this.  The decision to build depends on questions of scale, time and threat model.   Scale might be a single host, a local network event or even an organized entity with hundreds of thousands of systems.  Each notion of scale describes (when taken with a temporal component) a problem space in cyber security.  One commonly used strategy for detection is through anomaly detection, with the explicit assumption that any malicious behavior is anomalous~\cite{GaTa07}.  As a result, many approaches for anomaly detection have been proposed including,  support vector regression~\cite{Mirza2007}, $k$-means~\cite{Lakhina2005}, multivariate adaptive regression splines~\cite{Mukkamala2004}, Kalman filters~\cite{Soule2005}, and sequential hypothesis testing~\cite{Jung2004}.  As the data set sizes increase and the need to more quickly detect intrusions, more robust, accurate, and efficient algorithms will almost certainly be required.

\subsection{Attack Detection}
Current state of the art in attack detection provides many areas that can be assisted by improvements to model or algorithm design.  Network intrusion detection has traditionally been focused on identifying attackers who are seeking information about internal systems and services, sometimes over large address spaces or large time periods.  A fundamental component of this is {\em scan detection} where one or more remote network systems look over address ranges to survey available services.  To do this an adversary needs to scan some range of address space.  Single host detection has proven to be predictably accurate via sequential probability testing by Jung et al.~\cite{Jung2004}, but there are significant areas for improvement in the detection of distributed scanning, see for example Gates~\cite{Gates2009}.

Understanding how to represent both attack and defense is essential for developing a workable strategy.  Examples of this that might be extendable via a more complete analysis of the problem space are~\cite{Collins2008,Modelo-Howard2008}, which look at modeling an intrusion detection system's observable attack space as well as optimal placement strategies.  For example, Modelo-Howard~\cite{Modelo-Howard2008} proposed a new method based on a Bayesian network model that can characterize the relationship between attack steps and detectors.  This resulted in an algorithm that could evaluate the effect of detector configuration on system security.  A question that was left for future work was whether the solution was scalable to larger attack graphs and more detectors.  Similarly, Collins~\cite{Collins2008} argues that attacks should be viewed as a design specification, where the attacker is an engineer with specified goals.  His proposed solution involves estimating a detection surface via multiple Monte Carlo runs to build up a model for the probability of detection.  This naturally leads to the question of which models work best for different attack scenarios and how to best estimate them so as to reduce the number of false positives while still detecting the true attacks.

\subsection{Automated Attack Response}
Recently, there has been considerable work on developing capabilities to auto detect attacks based on both network and system behavior in order to reduce the time between attack detection and response.  Autogeneration of network signatures based on protocol and attack heuristics has been explored by Yegneswaran et al.~\cite{Yegneswaran2005.1,Yegneswaran2005.2}.  System call deviations based on static and dynamic analysis of downloaded binaries have also been studied by a number of people including Christodorescu et al.~\cite{Christodorescu2005}.

\subsection{Complex Systems and Other Novel Approaches}

Other types of models have recently been proposed. For example, Forrest and Hofmeyr~\cite{Forrest2000,Hofmeyr1999} have described models for network intrusion detection and virus detection based on an immunological distinction between ``selfÓ and ``nonself."  Using the analogy between an immune system they have studied problems in computer virus detection, host-based intrusion detection, automated response, and network intrusion detection.  For the network intrusion detection study, the system was tested on two months of network traffic data collected on a subnet of 50 computers at the Computer Science department at the University of New Mexico.  While preliminary, the results seem promising in that the false positive rate was on the order of two per day and the system was also able to successfully detect all seven intrusion incidents that were inserted into the system.

Another interesting approach was proposed by Zou et al.~\cite{Zou2005}. They proposed an adaptive defense principle based on minimizing a particular cost function.  The cost function depended on the attach severity, attack traffic and some other metrics that are determined by the types of attacks. They also presented a system design based on this approach to defend against SYN flood DDos attack and Internet worm infection.  

Finally, Zhou et al.~\cite{Zhou2007} have proposed several novel alert correlation algorithms for network intrusion detection that reduce the number of  false alerts.The basic building block of the model is a logical formula called a capability. They use the notion of capability to abstract consistently and precisely all levels of accesses obtained by the attacker in each step of a multistage intrusion.  The correlation algorithm is based on a new set searching algorithm that captures the case where multiple earlier attacks together support a new attack.  The experimental results of the correlator using several intrusion datasets demonstrate that the approach is effective in both alert fusion and alert correlation and has the ability to correlate alerts of complex multistage intrusions.

\section{Conclusions}
In this paper, we presented some of what we believe are the most important problems in cyber security for open science environments and highlighted those areas where mathematics and statistics could provide new approaches and solutions.  The use of mathematics and statistics in this field is relatively new and much remains to be done.  We also believe that the type of mathematics needed to address problems in cyber security will likely come from the use of non-traditional methods or techniques.  In summary, to paraphrase from Willinger and Paxson~\cite{WiPa98}, we believe that the field of  cyber security poses a rich set of new and exciting research opportunities for the mathematical and statistical sciences.

\bibliographystyle{plain}
\bibliography{CyberSecurity}

\begin{thebibliography}{10}

\bibitem{AlPaTe07}
Mark Allman, Vern Paxson, and Jeff Terrell.
\newblock A brief history of scanning.
\newblock In {\em IMC '07: Proceedings of the 7th ACM SIGCOMM conference on
  Internet measurement}, pages 77--82, New York, NY, USA, 2007. ACM.

\bibitem{Bishopet06}
M.~Bishop, R.~Crawford, B.~Bhumiratana, L.~Clark, and K.~Levitt.
\newblock Some problems in sanitizing network data.
\newblock In {\em 15th IEEE International Workshops on Enabling Technologies:
  Infrastructure for Collaborative Enterprise}, pages 307--312, 2006.

\bibitem{Bishop03}
Matt Bishop.
\newblock {\em Computer Security Art and Science}.
\newblock Addison Wesley, 2003.

\bibitem{Braun2008}
Uri Braun, Avraham Shinnar, and Margo Seltzer.
\newblock Securing provenance.
\newblock In {\em HOTSEC'08: Proceedings of the 3rd conference on Hot topics in
  security}, pages 1--5, Berkeley, CA, USA, 2008. USENIX Association.

\bibitem{Christodorescu2005}
M.~Christodorescu, S.~Seshia, S.~Jha, D.~Song, and R.~E. Bryant.
\newblock Semantics-aware malware detection.
\newblock In {\em IEEE Symposium on Security and Privacy}, June 2005.

\bibitem{Cleveland2000}
W.S. Cleveland and Don~X. Sun.
\newblock Internet traffic data.
\newblock {\em Journal American Statistical Association}, 95:979--985, 2000.
\newblock Reprinted in Statistics in the 21st Century , edited by A. E.
  Raftery, M. A. Tanner, and M. T. Wells, Chapman \& Hall/CRC, New York, 2002.

\bibitem{Collins2008}
M.~Collins and M.~Reiter.
\newblock On the limits of payload-oblivious network attack detection.
\newblock In {\em Recent Advances in Intrusion Detection}, May 2008.

\bibitem{Dreger2004}
Holger Dreger, Anja Feldmann, Vern Paxson, and Robin Sommer.
\newblock Operational experiences with high-volume network intrusion detection.
\newblock In {\em CCS '04: Proceedings of the 11th ACM conference on Computer
  and communications security}, pages 2--11, New York, NY, USA, 2004. ACM.

\bibitem{Dunlavy2009}
Daniel~M. Dunlavy, Bruce Hendrickson, and Tamara~G. Kolda.
\newblock Mathematical challenges in cybersecurity.
\newblock Technical Report SAND2009-0805, Sandia National Laboratories,
  February 2009.

\bibitem{Catlett2008}
Charlie~Catlett (Ed.).
\newblock A scientific research and development approach to cyber security.
\newblock Report submitted to the U.S. Department of Energy, 2008.

\bibitem{Floyd2001}
S.~Floyd and V.~Paxson.
\newblock Difficulties in simulating the internet.
\newblock {\em IEEE/ACM Transactions on Networking}, 9(4):392--403, Aug. 2001.

\bibitem{Forrest2000}
Stephanie Forrest and Steven~A. Hofmeyr.
\newblock Immunology as information processing.
\newblock In {\em Design Principles for the Immune System and Other Distributed
  Autonomous Systems, edited by L.A. Segel and I. Cohen. Santa Fe Institute
  Studies in the Sciences of Complexity}, pages 361--387. Oxford University
  Press, 2000.

\bibitem{GaTa07}
Carrie Gates and Carol Taylor.
\newblock Challenging the anomaly detection paradigm: a provocative discussion.
\newblock In {\em NSPW '06: Proceedings of the 2006 workshop on new security
  paradigms}, pages 21--29, New York, NY, USA, 2007. ACM.

\bibitem{Gates2009}
Carrrie Gates.
\newblock Coordinated scan detection.
\newblock In {\em 16th Annual Network and Distributed System Security
  Symposium}, San Diego, CA, February 8--11 2009. Internet Society.

\bibitem{NAPCyber}
Seymour~E. Goodman and Herbert~S. Lin, editors.
\newblock {\em Toward a Safer and More Secure Cyberspace}.
\newblock National Academies Press, Washington, DC, 2007.
\newblock Committee on Improving Cybersecurity Research in the United States,
  Computer Science and Telecommunications Board.

\bibitem{Hofmeyr1999}
Steve Hofmeyr and S.~Forrest.
\newblock Immunity by design: An artificial immune system.
\newblock In {\em Proceedings of the Genetic and Evolutionary Computation
  Conference}, San Francisco, CA, 1999. Morgan-Kaufmann.

\bibitem{Jung2004}
Jaeyeon Jung, Vern Paxson, Arthur~W. Berger, and Hari Balakrishnan.
\newblock Fast portscan detection using sequential hypothesis testing.
\newblock In {\em Proc. IEEE Symposium on Security and Privacy}, pages
  211--225, 9--12 May 2004.

\bibitem{WTE96}
F.~P. Kelly, S.~Zachary, I.~Ziedins, Walter Willinger, Murad~S. Taqqu, and
  Ashok Erramilli.
\newblock {\em A Bibliographical Guide to Self-Similar Traffic and Performance
  Modeling for Modern High-Speed Networks}, pages 339--366.
\newblock Clarendon Press, 1996.

\bibitem{Lakhina2005}
Anukool Lakhina, Mark Crovella, and Christophe Diot.
\newblock Mining anomalies using traffic feature distributions.
\newblock In {\em Proceedings {SIGCOMM}'05}, Philadelphia, PA, August 22--26
  2005. ACM.

\bibitem{Lakkaraju2008}
Kiran Lakkaraju and Adam Slagell.
\newblock Evaluating the utility of anonymized network traces for intrusion
  detection.
\newblock In {\em Proceedings SECURECOMM 2008}, Istanbul, Turkey, September
  22-25 2008.

\bibitem{Leland1994}
Will~E. Leland, Murad~S. Taqqu, Walter Willinger, and Daniel~V. Wilson.
\newblock On the self-similar nature of ethernet traffic (extended version).
\newblock {\em IEEE/ACM Trans. Netw.}, 2(1):1--15, 1994.

\bibitem{Markoff2009}
John Markoff.
\newblock Computer experts unite to hunt worm.
\newblock The New York Times, March 19, 2009.

\bibitem{McHugh2000}
John McHugh.
\newblock Testing intrusion detection systems: A critique of the 1998 and 1999
  {DARPA} off-line intrusion detection system evaluation as performed by
  {L}incoln {L}aboratory.
\newblock {\em ACO Transactions on Information and System Security},
  3(4):262--294, Nov. 2000.

\bibitem{Mirza2007}
Mariyam Mirza, Joel Sommers, Paul Barford, and Xiaojin Zhu.
\newblock A machine learning approach to tcp throughput prediction.
\newblock In {\em Proceedings {SIGMETRICS}'07}, San Diego, CA, June 2007. ACM.

\bibitem{Modelo-Howard2008}
G.~Modelo-Howard, S.~Bagchi, and G.~Lebanon.
\newblock Determining placement of intrusion detectors for a distributed
  application through bayesian network modeling.
\newblock In {\em Recent Advances in Intrusion Detection}, May 2008.

\bibitem{Mukkamala2004}
Srinivas Mukkamala, Andrew~H. Sung, Ajith Abraham, and Vitorino Ramos.
\newblock Intrusion detection systems using adaptive regression splines.
\newblock In {\em ICEIS (3)}, pages 26--33, 2004.

\bibitem{Paxson1995}
V.~Paxson and S.~Floyd.
\newblock Wide area traffic: the failure of poisson modeling.
\newblock {\em IEEE/ACM Transactions on Networking}, 3(3):226--244, June 1995.

\bibitem{Paxson1997a}
V.~Paxson and S.~Floyd.
\newblock Why we don't know how to simulate the internet.
\newblock In {\em Proc. Winter Simulation Conference}, pages 1037--1044, 7--10
  December 1997.

\bibitem{Paxson98}
Vern Paxson.
\newblock Bro: A system for detecting network intruders in real time.
\newblock In {\em Proceedings of the 7th USENIX Security Symposium}, Jan. 1998.

\bibitem{Paxson2004}
Vern Paxson.
\newblock Strategies for sound internet measurement.
\newblock In {\em Proceedings ACM IMC}, October 2004.

\bibitem{Raftery2001}
Adrian~E. Raftery, Martin~A. Tanner, and Martin~T. Wells, editors.
\newblock {\em Statistics in the 21st Century (Monographs on Statistics and
  Applied Probability)}.
\newblock Chapman and Hall/CRC, 2001.

\bibitem{Resnick97}
S.I. Resnick.
\newblock Heavy tail modeling and teletraffic data.
\newblock {\em The Annals of Statistics}, 25(5):1805--1869, 1997.

\bibitem{Slagell2006}
Adam~J. Slagell, Kiran Lakkaraju, and Katherine Luo.
\newblock Flaim: A multi-level anonymization framework for computer and network
  logs.
\newblock In {\em LISA}, pages 63--77, 2006.

\bibitem{Soule2005}
Augustin Soule, Kave Salamatian, and Nina Taft.
\newblock Combining filtering and statistical methods for anomaly detection.
\newblock In {\em Proceedings Internet Measurement Conference}, 2005.

\bibitem{Willinger1997}
Walter Willinger and Vern Paxson.
\newblock Discussion of {``Heavy Tail Modeling and Teletraffic Data"} by {S.I.}
  {R}esnick.
\newblock http://www.icir.org/vern/papers.html, 1997.

\bibitem{WiPa98}
Walter Willinger and Vern Paxson.
\newblock Where mathematics meets the internet.
\newblock {\em Notices of the American Mathematical Society}, 45:961--970,
  1998.

\bibitem{Willinger1997a}
Walter Willinger, Murad~S. Taqqu, Robert Sherman, and Daniel~V. Wilson.
\newblock Self-similarity through high-variability: statistical analysis of
  {E}thernet {LAN} traffic at the source level.
\newblock {\em IEEE/ACM Transactions on Networking}, 5:71--86, 1997.

\bibitem{Yegneswaran2005.1}
V.~Yegneswaran, P.~Barford, and V.~Paxson.
\newblock Using honeynets for internet situational awareness.
\newblock In {\em Proceedings of the ACM/USENIX Fourth Workshop on Hot Topics
  in Networks (Hotnets IV)}, Sept. 2005.

\bibitem{Yegneswaran2005.2}
V.~Yegneswaran, J.~Giffin, P.~Barford, and S.~Jha.
\newblock An architecture for generating semantics-aware signatures.
\newblock In {\em Proceedings of USENIX Security Symposium}, Sept. 2005.

\bibitem{Zhou2007}
Jingmin Zhou, Mark Heckman, Brennen Reynolds, Adam Carlson, and Matt Bishop.
\newblock Modeling network intrusion detection alerts for correlation.
\newblock {\em ACM Trans. Inf. Syst. Secur.}, 10(1):4, 2007.

\bibitem{Zou2005}
Cliff~C. Zou, Nick Duffield, Don Towsley, and Weibo Gong.
\newblock Adaptive defense against various network attacks.
\newblock In {\em SRUTI '05: Steps to Reducing Unwanted Traffic on the Internet
  Workshop}, 2005.

\end{thebibliography}

\end{document}